\documentclass[aps,prd,showpacs]{revtex4}
\usepackage[dvips]{graphicx}

\begin{document}

\title{Photons emerging as Goldstone bosons from spontaneous Lorentz
symmetry breaking: The Abelian Nambu model \\
}
\author{C. A. Escobar and L. F. Urrutia }
\affiliation{ Instituto de Ciencias Nucleares, Universidad Nacional Aut{\'o}noma de
M{\'e}xico, A. Postal 70-543, 04510 M{\'e}xico D.F., M{\'e}xico }

\begin{abstract}
After imposing current conservation together with the Gauss law as initial conditions
on the  Abelian Nambu model, we prove that the resulting theory is equivalent to standard QED in the non-linear gauge
$\left(A_{\mu }A^{\mu }-n^{2}M^{2}\right)=0$, to all orders in perturbation
theory. We show this by writing both models in terms of the same variables,
which produce identical Feynman rules for the interactions and propagators.
A crucial point is to verify that the Faddeev-Popov ghosts arising from the
gauge fixing procedure in the QED sector decouple to all orders. We verify
this decoupling by following a method like that employed in Yang-Mills
theories when investigating the behavior of axial gauges. The equivalence between the two theories
supports the idea that gauge particles can be envisaged as the Goldstone
bosons originating from spontaneous Lorentz symmetry breaking.
\end{abstract}

\maketitle

\section{Introduction}

The Abelian Nambu model (ANM) was proposed in Ref. \cite{Nambu-Progr} to
describe electrodynamics in a way similar to the construction of pion
interactions in the nonlinear sigma model characterized by spontaneous
chiral symmetry breaking. The understanding of pions as the Goldstone bosons
(GB) arising from such breaking motivated the possibility of looking at
photons as the GB resulting from a spontaneous Lorentz symmetry breaking
(SLSB). The ANM is defined by the Lagrangian density
\begin{equation}
\mathcal{L}=-\frac{1}{4}F_{\mu \nu }F^{\mu \nu }-J_{\mu }A^{\mu },\;\;
\label{LANM1}
\end{equation}%
plus the non-linear constraint
\begin{equation}
A_{\mu }A^{\mu }-n^{2}M^{2}=0,\;\;\;\;\;\;\;\;M>0.  \label{NLC}
\end{equation}%
The vector $n^{\mu }$ signals the direction of the non-zero vacuum
expectation value $\langle A^{\mu }\rangle =n^{\mu }M$ inducing the SLSB.
Usually, one deals separately with the three characteristic cases dictated
by the choice of the vector $n^{\mu }$ as time-like $(n^{2}>0)$,
space-like $(n^{2}<0)$ or light-like $(n^{2}=0)$. Here we consider only the
first two options where a natural choice of independent degrees of freedom
(DOF) is the following%
\begin{eqnarray}
n^{2} &>&0:\quad A_{1},\;A_{2},\;A_{3},\qquad \rightarrow \qquad A_{0}=\sqrt{%
M^{2}+A_{i}A_{i}},\qquad i=1,2,3,  \label{TL} \\
n^{2} &<&0:\quad A_{0},\;A_{1},\;A_{2},\qquad \rightarrow \qquad A_{3}=\sqrt{%
M^{2}+A_{0}A_{0}-A_{a}A_{a}},\qquad a=1,2.  \label{SL}
\end{eqnarray}%
The solutions of the constraint (\ref{NLC}) shown in Eqs. (\ref{TL}) and (%
\ref{SL}) make clear the symmetry breaking from $SO(1,3)$ to $SO(3)$ and
from $SO(1,3)$ to $SO(2,1)$ in the time-like and space-like cases,
respectively. Each realization of the ANM is defined by substituting the
dependent variable into the standard Lagrangian density for electrodynamics (\ref%
{LANM1}). The calculation of some particular processes in perturbation
theory demand a further expansion of the non linear terms in the resulting
Lagrangian density in powers of the combinations $\left( A_{i}A_{i}\right) /M^{2}$
(time-like case) and $\left( A_{0}A_{0}-A_{a}A_{a}\right) /M^{2}$
(space-like case). After the solutions (\ref{TL}) and (\ref{SL}) of the
constraint are inserted in the Lagrangian density (\ref{LANM1}) one ends up with a
regular system having three DOF per point in coordinate space, where the
resulting equations of motion are different from Maxwell's equations \cite%
{Nambu-Progr,Urru-Mont}.

Nambu models have been further studied in relation to electrodynamics \cite%
{Urru-Mont,VENT1,Azatov-Chkareuli} and also have been generalized to the
non-abelian \cite{VENT2,JLCH1,JLCH2,JLCH3} and gravitational cases \cite{JLCH4}. Quantum electrodynamics in the
nonlinear gauge (\ref{NLC}) is  considered in Refs. \cite{VENT1}. 
General conditions on how the gauge symmetries are recovered from
models that involve SLSB are worked out in Refs. \cite{RUSOS, JLCH5,JLCH6,JLCH7}. In
particular, the result obtained in Refs. \cite{Nambu-Progr,Urru-Mont,
NANM_ES_UR} states that the dynamics of the ANM guarantees the validity of
the Gauss Law for all times, once the Gauss law and current conservation are
imposed as initial conditions. In this way, after demanding such initial conditions, the ANM reduces to
electrodynamics and current conservation remain valid for all times as the
consequence of the restored gauge invariance.

An alternative way of exploring the connection between gauge theories and
models with SLSB is by means of the so-called bumblebee models appearing in the
study of possible observable violations of Lorentz invariance \cite%
{Colladay-Kostelecky}. These models introduce GB modes and depending on the
explicit form of the theory they present additional massive modes and
constraints. Such models have been thoroughly investigated in relation to
electrodynamics \cite{BLUHM_1,HERNASKI} and gravity \cite%
{BLUHM_3,BLUHM_2,KOS_POT,CARROLL}.

In this work we consider the relation between the ANM and standard QED from
a perturbative perspective, paying attention to the gauge fixing procedure
that is required in QED to study their equivalence. Perturbative
calculations in the ANM show that, to the order considered (tree level and
one-loop diagrams), all SLSB contributions to physical processes cancel out,
yielding the same results as in standard QED \cite%
{Nambu-Progr,Azatov-Chkareuli}. This feature has been interpreted by stating
that the non-linear constraint (\ref{NLC}) can be seen just as a
gauge choice in QED, which would then explain why the two theories are
equal. Nevertheless, this statement requires some qualifications: (i) on one
hand, the number of degrees of freedom (DOF) of the ANM is three, in such a
way that the possible equivalence between both theories requires at least to
specify some additional condition to cut this extra DOF. (ii) on the other
hand, fixing the gauge in any gauge theory requires the introduction of
ghost particles, via the BRST procedure  for example, which play a fundamental role as internal particles in calculating
physical processes \cite%
{BRST}. Thus, to show the proposed equivalence one would need to
study their contributions. A possible decoupling of such ghosts is by no
means clear, especially due to the non-linear character of the suggested
gauge fixing.

The paper is organized as follows. In section 2 we review some basic points
of the perturbative calculation presented in Ref. \cite{Azatov-Chkareuli}, where the authors
introduce a convenient field redefinition $A_{\mu }\rightarrow a_{\mu }$, which allows to write the ANM in terms of the GB modes only.
This formulation serves as the benchmark for the
comparison of the ANM with QED. Section 3 describes  how the BRST
formalism fixes the gauge $A_{\mu }A^{\mu }=n^{2}M^{2}$ in QED, introducing
the required Faddeev-Popov ghosts (FPG). The resulting gauge fixed QED
Lagrangian density is then written in terms of the same field redefinition
$A_{\mu }\rightarrow a_{\mu }$ already introduced in section 2. In this way,
one can show that the Feynman amplitudes for  physical processes arising from the
ANM Lagrangian density and those stemming form the gauge fixed QED Lagrangian density differs
only by the contributions from the Feynman diagrams including the FPG
interactions. The ghost propagator, together with the other Feynman rules,
is calculated in section 4. Finally, section 5 includes the general
perturbative proof on how the ghosts decouple when QED is formulated in the
gauge fixed Lagrangian density  found in section 3. In section 6 we close with a
summary, where we put together all the pieces which prove the perturbative
equivalence between the two models.

\section{The perturbative formulation of the ANM}

In this section, we summarize the approach of Ref. \cite{Azatov-Chkareuli},
which is appropriate to make explicit the relation between the ANM and QED
to be established in the following.

The starting point in Ref. \cite{Azatov-Chkareuli} is the standard fermionic
QED Lagrangian density
\begin{equation}
\mathcal{L}_{\mathrm{QED}}(A_{\mu },{\bar{\psi}},\psi )=\mathcal{-}\frac{1}{4%
}F_{\mu \nu }F^{\mu \nu }+\bar{\psi}\left( i\gamma \partial +m\right) \psi
-eA_{\mu }\bar{\psi}\gamma ^{\mu }\psi ,  \label{LSQED}
\end{equation}%
plus the constraint (\ref{NLC}), where the authors then introduce a very useful parameterization\ in terms of
the new field $a_{\mu }$, by defining%
\begin{equation}
A_{\mu }(a_{\rho })=a_{\mu }+\frac{n_{\mu }}{n^{2}}(M^{2}-n^{2}a^{2})^{1/2},%
\quad \;n^{2}\neq 0.  \label{CHANGV}
\end{equation}%
This transformation can be inverted yielding
\begin{equation}
a_{\mu }=A_{\mu }-\frac{n_{\mu }}{2n^{2}}\left( \left( n\cdot A\right) +%
\sqrt{\left( n\cdot A\right) ^{2}+2\left( M^{2}-A^{2}n^{2}\right) }\right) .
\label{INVCHANGV}
\end{equation}
When we substitute (\ref{CHANGV}) into (\ref{LSQED}) we get $\mathcal{L}_{%
\mathrm{QED}}(A_{\mu }(a_{\nu }),{\bar{\psi}},\psi )$, which is a highly
non-linear expression in terms of the new field$\;a_{\mu }.\;$Nevertheless,$%
\;\mathcal{L}_{\mathrm{QED}}(A_{\mu }(a_{\nu }),{\bar{\psi}},\psi )\;$is
still a Lagrangian density  for QED, written in a very unconventional way, which
nevertheless provides a convenient interpretation of the field $a_{\mu }$\
once the ANM is defined. Notice that gauge invariance $\delta A_{\mu
}=\partial _{\mu }\delta \Lambda \;$remains an invariance of $\;\mathcal{L}_{%
\mathrm{QED}}(A_{\mu }(a_{\nu }),{\bar{\psi}},\psi )\;$realized in terms of
a very complicated transformation $\delta a_{\mu }$,\ which could be
obtained from Eq. \ (\ref{INVCHANGV}). An important property of the field
redefinition (\ref{CHANGV}) is the relation
\begin{equation}
\left( A_{\mu }A^{\mu }-n^{2}M^{2}\right) ^{2}=4\left( a\cdot n\right) ^{2}
\left[ M^{2}-n^{2}a^{2}\right] .  \label{ANMC2}
\end{equation}
Next, the authors of Ref. \cite{Azatov-Chkareuli} focus on the ANM, by
imposing the non-linear constraint (\ref{NLC}) on the Lagrangian density $\mathcal{L}%
_{\mathrm{QED}}(A_{\mu }(a_{\nu }),{\bar{\psi}},\psi )$. In terms of the new
fields $a_{\nu }$, the condition (\ref{NLC})\ takes the simpler form
\begin{equation}
n\cdot a=0,  \label{NLC1}
\end{equation}%
according to the relation (\ref{ANMC2}). The fields $a_{\nu }$, satisfying (%
\ref{NLC1}), define three DOF that are orthogonal to the vacuum $n^{\mu },$
so that they describe the Goldstone modes of the model.

Since the ANM is defined by solving a constraint involving four fields in
terms of three DOF and after substituting the solution in the
electromagnetic sector of the model, it is enlightening to compare the two
possibilities offered once we introduce the field $a_{\nu }$, to appreciate
the advantage of this redefinition. For the time-like case $n^{\mu
}=(1,0,0,0)$, the solution $A_{0}=\sqrt{M^{2}+A_{i}A_{i}}$, together with
the choice of $A_{i}$ as the independent variables is fully equivalent to
set $a_{0}=0$ and recognize that $a_{i}=A_{i}$. For the space-like case $%
n^{\mu }=(0,0,0,1)$ the choice of independent variables $A_{0},A_{a}$,
together with the definition $A_{3}=\sqrt{M^{2}+A_{0}A_{0}-A_{a}A_{a}}$,
corresponds to set $a_{3}=0,a_{0}=A_{0}\;$and $a_{a}=A_{a}$. Thus, the
substitution of $A_{0}$ or $A_{3}$ into (\ref{LSQED}), according to the
choices (\ref{TL}) or (\ref{SL}), is equal to the introduction of the field
redefinition (\ref{CHANGV}) with the proper choice of $n^{\mu }$, which we
can select at the end of the calculation. This provides a unified method of
dealing with the time-like and space-like cases. Consequently, and following
Ref. \cite{Azatov-Chkareuli}, we can write the Lagrangian density for the cases of
interest in the ANM as
\begin{equation}
\mathcal{L}_{\mathrm{ANM}}(a_{\mu },{\bar{\psi}},\psi )=\mathcal{L}_{\mathrm{%
QED}}(A_{\mu }(a_{\nu }),{\bar{\psi}},\psi ),\;\;\;n\cdot a=0.\;
\label{LANM_LQED}
\end{equation}%
The constraint (\ref{NLC1}) clearly breaks gauge invariance together with
active Lorentz invariance. The field redefinition $A_{\mu }(a_{\rho })$
implies that we have to substitute
\begin{equation}
F_{\mu \nu }=f_{\mu \nu }+\frac{1}{n^{2}}\left( n_{\nu }\partial _{\mu
}-n_{\mu }\partial _{\nu }\right) (M^{2}-n^{2}a^{2})^{1/2},\;\;\;f_{\mu \nu
}=\partial _{\mu }a_\nu -\partial _{\nu }a_{\mu },\;\;\;  \label{F(a)}
\end{equation}
for the field strength $F_{\mu\nu}$ in (\ref{LSQED}).

The next step in Ref. \cite{Azatov-Chkareuli} is to make an expansion of $%
\mathcal{L}_{\mathrm{QED}}(A_{\mu }(a_{\nu }),{\bar{\psi}},\psi )$ in powers
of $a^{2}/M^{2}$ keeping terms up to the order $(a^{2})^{2}/M^{2}$, which
defines the Lagrangian density of the ANM to the order considered in that reference.
The corresponding Feynman rules are given in Ref.\ \cite{Azatov-Chkareuli}
and we recall here the GB propagator $D_{\mu \nu }^{\mathrm{ANM}}(k)$ which
we will need in the following
\begin{equation}
D_{\mu \nu }^{\mathrm{ANM}}(k)=-\frac{i}{k^{2}+i\epsilon }\left[ \eta _{\mu
\nu }-\frac{n_{\mu }k_{\nu }+n_{\nu }k_{\mu }}{\left( n\cdot k\right) }+%
\frac{n^{2}k_{\mu }k_{\nu }}{\left( n\cdot k\right) ^{2}}\right]
,\;\;\;n^{\mu }D_{\mu \nu }^{\mathrm{ANM}}=0\;.  \label{FOTPROPANM}
\end{equation}

We remind the reader that the authors of Refs. \cite%
{Nambu-Progr,Azatov-Chkareuli} have shown that the extra contributions to
some specific QED processes (up to one loop order), arising from the Lorentz
violating terms in the Lagrangian density (\ref{LANM_LQED}) exactly cancel in the
ANM calculation, thus yielding the standard QED results.

Let us emphasize that the perturbative calculation naturally incorporates
the two initial conditions required for the equivalence between the ANM and
QED, which are the imposition of the Gauss law together with current
conservation \cite{Nambu-Progr,Urru-Mont, NANM_ES_UR}. The Lagrangian density of the
ANM in the interaction picture starts with the free contribution
\begin{equation}
\mathcal{L}_{0}(a_{\mu },{\bar{\psi}},\psi )=\mathcal{-}\frac{1}{4}f_{\mu
\nu }f^{\mu \nu }+\bar{\psi}\left( i\gamma \partial +m\right) \psi
,\;\;\;\;n\cdot a=0,\;  \label{LANMIP}
\end{equation}%
which describes the behavior of the system at $t\;\rightarrow -\infty$. The
electric current $J^{\mu }=e\bar{\psi}\gamma ^{\mu }\psi $ is conserved,
as it is  the Noether current arising from the invariance under global phase
transformations of the fermionic fields in the Lagrangian density (\ref{LANMIP}).
Also, the Lagrangian density (\ref{LANMIP}) yields the GB propagator (\ref{FOTPROPANM}),
which satisfies the on-shell
condition \ $k^{\mu }D_{\mu \nu }^{\mathrm{ANM}}=0$, for $k^{2}=0$. That is
to say, the Gauss law has been implemented \`{a} la Dirac upon the initial
physical states, by imposing the transversality condition $k^{\mu }\epsilon
_{\mu }(k)=0$ on the external GB modes having the polarization
vectors $\epsilon _{\mu }(k)$.  Both conditions play a crucial role in the
cancellations obtained in Ref. \cite{Azatov-Chkareuli}, which suggest the
equivalence, to this order in perturbation theory, between  QED and the ANM
with appropriate initial conditions.

\section{Electrodynamics in the gauge $A_{\protect\mu }A^{\protect\mu%
}=n^{2}M^{2}$}

Now we switch to electrodynamics. The main point we address in this section
is the behavior of the Faddeev-Popov ghost (FPG) interactions that will
necessarily appear when fixing the proposed gauge. We move forward after the
basic prescription of the BRST method \cite{BRST}, applied to the QED the
Lagrangian density (\ref{LSQED}), which is invariant under the gauge transformations
$\delta A_{\mu }=\partial _{\mu }\Lambda $. First we introduce the fermionic
nilpotent transformation $\tilde{\delta}$, together with the new fields $%
c,\;\bar{c}$ and $b$ in such a way that
\begin{equation}
\tilde{\delta}A_{\mu }=\partial _{\mu }c,\qquad \tilde{\delta}c=0,\qquad
\tilde{\delta}\bar{c}=ib,\qquad \tilde{\delta}b=0.
\end{equation}%
Next we construct the BRST invariant Lagrangian density
\begin{equation}
\mathcal{L}_{\mathrm{BRST}}=-\frac{1}{4}F_{\mu \nu }F^{\mu \nu }+i\tilde{%
\delta}\left[ \bar{c}\left( \left( A_{\mu }A^{\mu }-n^{2}M^{2}\right) +\frac{%
b}{2\alpha }\right) \right] ,
\end{equation}%
where we have explicitly introduced the gauge fixing condition. Performing
the $\tilde{\delta}$ variation and eliminating $b$ from its algebraic
equation of motion, we get the Lagrangian density%
\begin{equation}
\mathcal{L}_{\mathrm{BRST}}=-\frac{1}{4}F_{\mu \nu }F^{\mu \nu }-\frac{%
\alpha }{2}\left( A_{\mu }A^{\mu }-n^{2}M^{2}\right) ^{2}-2A^{\mu }i\bar{c}%
\partial _{\mu }c,
\end{equation}%
which clearly exhibits $\,\alpha \left( A_{\mu }A^{\mu }-n^{2}M^{2}\right)
^{2}/2\,$ as the gauge fixing term and brings in the FPG $c$ and $\bar{c}$,
which are independent real Grassman numbers. At this stage it is convenient
to introduce also the parameterization (\ref{CHANGV}) for the photon field $%
A^{\mu }$. Recalling that such parameterization yields the exact result (\ref%
{ANMC2}), the final gauge fixed Lagrangian density for QED is
\begin{eqnarray}
\mathcal{L}_{\mathrm{GFED}} &=&\mathcal{L}_{\mathrm{QED}}(A_{\mu }(a_{\nu }),%
{\bar{\psi}},\psi )\nonumber \\
&&-2\alpha M^{2}\left( n\cdot a\right) ^{2}+2\alpha
n^{2}a^{2}\left( n\cdot a\right) ^{2}  \nonumber \\
&&-2\left( a^{\mu }+\frac{n^{\mu }}{n^{2}}(M^{2}-n^{2}a^{2})^{1/2}\right) i%
\bar{c}\partial _{\mu }c,  \label{LAGFED1}
\end{eqnarray}%
where we have only added the gauge fixing contributions to the Lagrangian density $%
\mathcal{L}_{\mathrm{QED}}(A_{\mu }(a_{\nu }),{\bar{\psi}},\psi )$. Let us
emphasize that $\mathcal{L}_{\mathrm{QED}}(A_{\mu }(a_{\nu }),{\bar{\psi}}%
,\psi )$ appearing in Eq. (\ref{LAGFED1}) is the same Lagrangian density  which defines
 the ANM in Eq. (\ref{LANM_LQED}). Notice that the gauge fixing
term reduces to $2\alpha M^{2}\left( n\cdot a\right) ^{2}$, when written in
terms of the redefined photon field $a_{\mu }$. This corresponds to the
choice of an axial gauge in electrodynamics, but with added Yang-Mills type
interactions. That is to say, the field $a^{\mu }$ now describes photons in
the gauge $\left( n\cdot a\right) =0$, instead of GB modes as in the
previous section. From Ref. \cite{LEIBBRANDT} we read the photon propagator
in the axial gauge
\begin{equation}
D_{\mu \nu }^{\mathrm{QED}}(k)=\frac{-i}{k^{2}+i\epsilon }\left[ \eta _{\mu
\nu }-\frac{k_{\mu }n_{\nu }+n_{\mu }k_{\nu }}{\left( n\cdot k\right) }%
+k_{\mu }k_{\nu }\frac{n^{2}+\frac{k^{2}}{4\alpha M^{2}}}{\left( n\cdot
k\right) ^{2}}\right] ,  \label{PROPLEIB}
\end{equation}%
satisfying%
\begin{equation}
k^{\mu }D_{\mu \nu }^{\mathrm{QED}}(k)=0,\qquad n^{\mu }D_{\mu \nu }^{%
\mathrm{QED}}(k)=\frac{-i}{k^{2}+i\epsilon }\left[ \frac{k^{2}k_{\nu }}{%
4M^{2}\left( n\cdot k\right) }\right] \frac{1}{\alpha }.  \label{GAUGECOND}
\end{equation}%
Next we choose the so called pure (homogeneous) axial gauge, defined by $%
\alpha \rightarrow \infty $, and see that the propagator (\ref{PROPLEIB})
reduces to (\ref{FOTPROPANM}), that is precisely the one employed in the ANM
calculation of Ref. \cite{Azatov-Chkareuli}, which nevertheless arises in QED
from the gauge fixing.

The next step in dealing with the contribution to physical processes of the
terms in the QED gauge fixed Lagrangian density is to consider the two non-linear
terms in the second line of Eq. (\ref{LAGFED1}), which do not depend on the
FPG and which are not present in the ANM Lagrangian density of Eq. (\ref%
{LANM_LQED}). The last term in that line is proportional to $a^{2}\left(
n\cdot a\right) ^{2}$ and produces a four-photon vertex $V_{\alpha \beta \mu
\nu }$%
\begin{equation}
V_{\alpha \beta \mu \nu }\sim \alpha (\eta _{\alpha \beta }n_{\mu }n_{\nu
}+perm.\;).  \label{V4a}
\end{equation}%
When we saturate $V_{\alpha \beta \mu \nu }$ with internal photon lines, via
the corresponding propagators, this vertex will have two contractions of the
type $n^{\mu }D_{\mu \rho }$ which give a factor $1/\alpha ^{2}$. In this
way, the net contribution goes like $1/\alpha $ and thus cancels in the
limit $\alpha \rightarrow \infty $. On the other hand, for on-shell photons
we have $n^{\mu }D_{\mu \nu }(k)=0=k^{\mu }D_{\mu \nu }(k)$ in such a way
that their polarizations vectors \ $\epsilon _{\mu }(k)$ must satisfy $%
n^{\mu }\epsilon _{\mu }(k)=0=k^{\mu }\epsilon _{\mu }(k)$. These added
conditions lead also to a zero contribution when we attach external photons
to $V_{\alpha \beta \mu \nu }$. The same argument applies to the gauge
fixing term proportional to $\left( n\cdot a\right) ^{2}$. In other words,
the terms proportional to $\left( n\cdot a\right) $ just carry out the gauge
condition $\left( n\cdot a\right) =0$.

Then, in the pure axial gauge, the gauge fixed Lagrangian density ${\mathcal{L}_{\mathrm{GFED}}}$ for fermionic
electrodynamics, in the parameterization of Ref. \cite{Azatov-Chkareuli},
reads%
\begin{equation}
\mathcal{L}_{\mathrm{GFED}}=\mathcal{L}_{\mathrm{QED}}(A_{\mu }(a_{\nu }),{%
\bar{\psi}},\psi )+\mathcal{L}_{\mathrm{GHOST}},  \label{GFQEDL}
\end{equation}%
which we compare with the ANM Lagrangian density in Eq. (\ref{LANM_LQED}).
In both cases the requirement $\left( n\cdot a\right)=0$ holds, however for
different reasons: it is a defining condition in the ANM, while it corresponds to
a gauge fixing condition in QED. Let us recall that the field $a_{\mu }\;$%
has the same propagator in both theories, as seen from \ Eq. (\ref%
{FOTPROPANM}) for the ANM  and from the limit $\alpha \rightarrow \infty $
in Eq.\ (\ref{PROPLEIB}) for QED. In this way, the condition $\left( n\cdot
a\right)=0$ is implemented in the same way for each model in terms of the
Feynman rules for the $a_\mu$ propagator. This propagator also imposes the Gauss law as an initial
condition in the ANM by demanding transversality of the on-shell GB. Besides, as already mentioned, the perturbative expansion in terms of fields in
the interaction picture guarantees current conservation as an initial
condition in the ANM. Moreover, the remaining Feynman rules for the fields $%
a_{\mu },{\bar{\psi}},\psi$, which arise from $\mathcal{L}_{\mathrm{QED}%
}(A_{\mu }(a_{\nu }),{\bar{\psi}},\psi )$, are the same in both cases. In this way,
 the only difference between the perturbative expansions of the
ANM and the gauge fixed QED arises from the ghost interactions%
\begin{equation}
\mathcal{L}_{\mathrm{GHOST}}=-2\left( a^{\mu }+\frac{Mn^{\mu }}{n^{2}}\left(
1-\frac{n^{2}a^{2}}{M^{2}}\right) ^{1/2}\right) i\bar{c}\partial _{\mu }c,
\label{LGHOST}
\end{equation}%
which contributions we study in the following sections.

\section{The Feynman rules for Electrodynamics in the gauge $A_{\protect\mu %
}A^{\protect\mu }=n^{2}M^{2}$}

In this section we find the extra Feynman rules arising from the FPG
couplings to the photon in the gauge-fixed QED. Since $\bar{c}$ and $c$ are
two independent real Grassman fields we find it more convenient to introduce
the real doublet
\begin{equation}
\left[
\begin{array}{c}
c \\
\bar{c}\;%
\end{array}%
\right] =\frac{1}{\sqrt{2}}\Phi ,\qquad \Phi =\Phi ^{\ast },
\end{equation}%
in terms of which we can write
\begin{equation}
i\bar{c}\partial _{\mu }c=\frac{i}{4}\Phi \sigma ^{(1)}\partial _{\mu }\Phi -%
\frac{i}{2}\partial _{\mu }(c{\bar{c}}),\qquad \sigma ^{(1)}=\left[
\begin{array}{cc}
0 & 1 \\
1 & 0%
\end{array}%
\right] =(\sigma ^{(1)})^{T}.  \label{GBTERM}
\end{equation}%
To get the ghost Feynman rules we need to expand Eq. (\ref{LGHOST}) in
powers of $n^{2}a^{2}/M^{2}$, in the same way as done for the ANM. We get
\begin{eqnarray}
&&\hskip-.7cm\mathcal{L}_{\mathrm{GHOST}}=-\frac{Mn^{2}}{2}\Phi \sigma
^{(1)}\left( n\cdot i\partial \right) \Phi -a^{\mu }\Phi \sigma
^{(-)}i\partial _{\mu }\Phi +  \nonumber \\
&&\hskip-.7cm+\left( \Phi \sigma ^{(-)}\left( n\cdot i\partial \right) \Phi
\right) \sum_{m=1}\left( \frac{(2m-3)!!}{2^{m}m!}\right) \left( \frac{\left(
(n^{\mu }n_{\mu })\right) ^{m-1}}{M^{2m-1}}\right) \left( a^{2}\right) ^{m},
\label{FGHOST}
\end{eqnarray}%
where we have substituted Eq. (\ref{GBTERM}) and omitted a total derivative.
We have rewritten the ghost-photon  interaction in terms of the field $\Phi $ by
introducing $\sigma ^{(-)}=(\sigma ^{(1)}-i\sigma ^{(2)})/2$, where $\sigma
^{(2)}$ is the standard Pauli matrix. We indicate the Feynman rules for the
FPG obtained from (\ref{FGHOST}) in the following figures. Dashed lines
denote the FPG and wavy lines denote the photon. Figure 1 shows the FPG
propagator, which arises from the kinetic term in Eq. (\ref{FGHOST}). Figure
2 shows the vertex $V^{1}$ describing the ghost-photon interaction. Figure 3
shows the vertices ${V}^{2m},m=1,2,\cdots $ describing the ghost-($2m$ photons)
interaction. The function $V_{\alpha _{1}\beta _{1}\alpha _{2}\beta
_{2}....\alpha _{m}\beta _{m}}$ is independent of the momenta and it is
given by
\begin{equation}
V_{\alpha _{1}\beta _{1}\alpha _{2}\beta _{2}....\alpha _{m}\beta _{m}}=%
\frac{\left( n^{2}\right) ^{m-1}(2m-3)!!}{M^{2m-1}}\left( \eta _{\alpha
_{1}\beta _{1}}\eta _{\alpha _{2}\beta _{2}}.....\eta _{\alpha _{m}\beta
_{m}}+perm\right).  \label{V2M}
\end{equation}%
\begin{figure}[th]
\centering \includegraphics[scale=0.5]{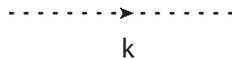}
\caption{The ghost propagator: $s(k)=\protect\sigma ^{(1)}\frac{i}{M}\frac{n^{2}}{(n\cdot k)}$.}
\label{figura1}
\end{figure}
The permutations in Eq.(\ref{V2M}) do not include repetitions and there are $%
(2m)!/(2^{m}m!)$ terms for a given $m$. We suppress tensor indices in the
notation $V^{1}$ and ${V}^{2m}$, which can be recovered from the
definitions in Figure 2 and Figure 3, together with Eq.(\ref{V2M}).
\begin{figure}[th]
\centering \includegraphics[scale=0.5]{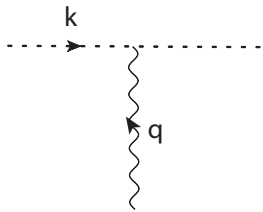}
\caption{The ghost-photon vertex: $V^{1}(k^{\mu} )=\protect\sigma ^{(-)}\frac{i}{2}k^{\mu} $.}
\label{figura2}
\end{figure}

\section{Faddeev-Popov Ghosts contributions to physical processes}

Finally, we consider the extra Feynman amplitudes to physical processes arising from
the FPG interactions. Since FPG couple only to the photon, their contribution
 will appear as internal loops which generic diagram is shown in
Figure 4.
\begin{figure}[th]
\centering \includegraphics[scale=0.5]{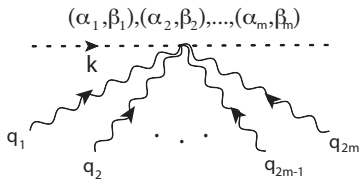}
\caption{The ghost-($2m$ photons) vertex: $V^{2m}(k^{\protect\mu })=-\protect\sigma ^{(-)}\frac{i}{2}(n\cdot
k)\,V_{\protect\alpha _{1}\protect\beta _{1}\protect\alpha _{2}\protect\beta %
_{2}\dots \protect\alpha _{m}\protect\beta _{m}}$.}
\label{figura3}
\end{figure}
This loop has $N$ vertex insertions which can be of  two types: one photon vertex
 $V^{1}(k)$ and $2m$-photons vertex  $V^{2m}(k)$. They
will be generically denoted by $V^{A}(k_{A}),\;\;A=1,2,4,\cdots ,2m,\cdots $%
, where $k_{A}$ indicates the ghost momentum coming into the corresponding
vertex. We keep track of the position of each vertex in the loop by an
additional subindex $n$ in $V_{n}^{A_{n}}(k_{A_{n}}),$ where $%
A_{n}=1,2,\cdots ,2m,\cdots$, labels the type of the vertex $n$ and $%
k_{A_{n}}$ stands for  the incoming ghost momentum.
\begin{figure}[th]
\centering \includegraphics[scale=0.5]{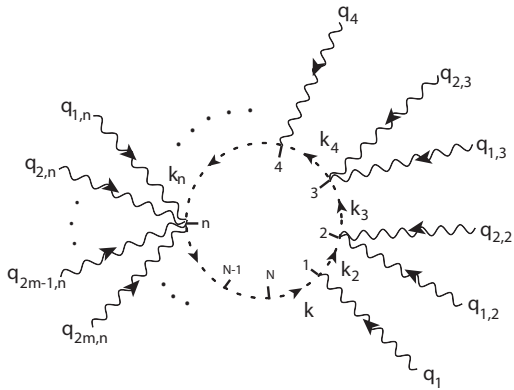}
\caption{General closed loop ghost diagram with N vertices, showing
arbitrary insertions of type $V^{1}$ and type $V^{2m}$. All photons are
incoming. The photons arriving to vertices $(N-1)$ and $N$ are not shown.}
\label{figura4}
\end{figure}
Each vertex $V^{A}(k_{A})$ has also incoming external photon momenta denoted
by $Q_{A}$. For a type $1$ vertex, $Q_{A}$ is just the momentum of the
incoming photon, while for a type $2m$ vertex in the $n$-place we have $%
Q_{A_{n}=2m}=[q_{1,n}+q_{2,n}+\cdots +q_{2m,n}]$, which is the sum of the
momenta of the $2m$ incoming photons. The basic unit forming the loop is the
product of the propagator \ $s_{n}^{A_{n}}(k_{A_{n}})\;\ $corresponding to
the ghost coming into each vertex, times the corresponding vertex factor$%
\;V_{n}^{A_{n}}(k_{A_{n}})$, which results in $s_{n}^{A_{n}}(k_{A_{n}})%
\times V_{n}^{A_{n}}(k_{A_{n}})$. Let us denote  $k_{A_{1}}=k$ the momentum entering the
first ($n=1$) vertex $V_{1}^{A_{1}}$, which will serve as
the integration variable over the loop. Without writing the external
connections of the photons entering each vertex and using dimensional
regularization, we find that a ghost loop with $N$ vertices contributes to
the amplitude with
\begin{eqnarray}
iM_{L} &\sim &\int \frac{d^{D}k}{(2\pi )^{D}}Tr\left( \left[
s_{1}^{A_{1}}(k_{A_{1}}=k)V_{1}^{A_{1}}(k_{A_{1}})\right] \left[
s_{2}^{A_{2}}(k_{A_{2}})V_{2}^{A_{2}}(k_{A_{2}})\right] \right.  \nonumber \\
&&\left. \times .......\times \left[
s_{n}^{A_{n}}(k_{A_{n}})V_{n}^{A_{n}}(k_{A_{n}})\right] ......\left[
s_{N}^{A_{N}}(k_{A_{N}})V_{N}^{A_{N}}(k_{A_{N}})\right] \right) .
\end{eqnarray}%
We have
\begin{equation}
k_{A_{\left( n+1\right) }}=Q_{A_{n}}+k_{A_{n}},\quad k_{A_{1}}=k,\quad
k_{A_{n}}=k+Q_{A_{1}}+\cdots +Q_{A_{\left( n-1\right) }}.
\end{equation}
Here, $Q_{A_{n}}$ are the total external photon momenta entering the loop
through the vertex $n$, \ satisfying$\;\;\sum_{n=1}^{N}Q_{A_{n}}=0$, so that
the ghost momentum after the vertex $N$\ is equal to\ $k=k_{A_{1}}$. The
loop has $N$ vertices and $N$ propagators, each contributing with a matrix
factor $\sigma ^{(-)}$ and $\sigma ^{(1)}$, respectively. In this way, the
trace corresponds to $Tr\left( (\sigma ^{(-)}\sigma ^{(1)})^{N}\right) =1$.

Some important simplifications occur due to the specific form of the ghost
propagator and the ghost-($2m$ photons) vertex. In fact, when the $r$ vertex is of
type $2m$, the dependence on the factor $n\cdot (k_{A_{r=2m}})$ in the
vertex cancels with that of the propagator, yielding the momentum
independent contribution
\begin{equation}
s_{r}^{A_{r}=2m}(k_{A_{r}=2m})V_{r}^{A_{r}=2m}(k_{A_{r}=2m})=\left( \frac{1}{%
2M^{2}}\right) V_{\alpha _{1}\beta _{1}\cdots \alpha _{m}\beta _{m}},
\end{equation}%
that can be taken out of the $d^{D}k$ integration and which is not written
in the following steps.

In this way, we are left with contributions to the diagram arising only from
vertices of type $1$, together with the respective propagators. Thus, we
have
\begin{eqnarray}
iM_{L} &\sim &\int \frac{d^{D}k}{(2\pi )^{D}}Tr\left( \Pi _{a=1}^{N_{1}}%
\left[ \left( \sigma ^{(1)}\frac{i}{M}\frac{n^{2}}{(n\cdot k_{a})}\right)
\left( \sigma ^{(-)}\frac{i}{2}k_{a}^{\mu _{a}}\right) \right] \right)
\nonumber \\
iM_{L} &\sim &\left( -\frac{n^{2}}{2M}\right) ^{N_{1}}\int \frac{d^{D}k}{%
(2\pi )^{D}}\frac{k_{1}^{\mu _{1}}k_{2}^{\mu _{2}}...k_{N_{1}}^{\mu _{N_{1}}}%
}{(n\cdot k_{1})(n\cdot k_{2})...(n\cdot k_{N_{1}})},  \label{LM3}
\end{eqnarray}%
where $a=1,\cdots ,N_{1}$ relabels the remaining $N_{1}$ type $1$ vertices.
Let us denote $k_{a}=k+R_{a}$, where $R_{a}$ is the sum of all the external
photon momenta that have entered the loop before the vertex $a$ and after
the vertex $1$. Using the standard Feynman's parameterization for the
denominators,%
\begin{equation}
\frac{1}{\left[ n\cdot (k+R_{1})\right] ...\left[ n\cdot (k+R_{N_{1}})\right]
}=\int_{0}^{1}d\alpha _{1}\cdots d\alpha _{N_{1}}\delta (\alpha _{1}+\cdots
+\alpha _{N_{1}}-1)\frac{(N_{1}-1)!}{\left[ n\cdot \left( k+V\left( \alpha
\right) \right) \right] ^{N_{1}}},
\end{equation}%
with $V\left( \alpha \right)=\sum_{a=1}^{N_{1}}\alpha _{a}R_{a}$, the
contribution from Eq.(\ref{LM3}) reduces to%
\begin{equation}
iM_{L}\sim \int \frac{d^{D}k}{(2\pi )^{D}}\left( k+S_{1}(\alpha )\right)
^{\mu _{1}}\left( k+S_{2}(\alpha )\right) ^{\mu _{2}}\left(
k+S_{N_{1}}(\alpha )\right) ^{\mu _{N_{1}}}\frac{1}{\left( n\cdot k\right)
^{N_{1}}}.  \label{FINALML}
\end{equation}%
Here $S_{a}(\alpha )=R_{a}-V\left( \alpha \right) $. Also, we have made  the shift $%
k+V\left( \alpha \right) \rightarrow k$ in the integration
variable of Eq. (\ref{LM3}). In Eq. (\ref{FINALML}) we have omitted the
integrations with respect to the Feynman parameters $\alpha _{a}$, which can
be taken out of the momentum integral. The numerator in the integral (\ref%
{FINALML}) is a linear combination of products of the type
\begin{equation}
k^{\nu _{1}}k^{\nu _{2}}\cdots k^{\nu _{L}},\qquad L=0,1,\cdots ,N_{1},
\end{equation}%
multiplied by $M$ constant vectors $S^{\rho _{i}}S^{\rho _{2}}...S^{\rho
_{M}}$,\ in such a way that $M+L=N_{1}$. The set of $L$ indices $\nu
_{1},\nu _{2},\cdots ,\nu _{L}$ is chosen among the $N_{1}$ original ones $%
\mu _{1},\mu _{2},\cdots ,\mu _{L},\cdots ,\mu _{N_{1}}$, in all possible
combinations. Finally, the momentum integral of the ghost loop contributes
with a sum of integrals of the form%
\begin{equation}
I^{{\nu _{1}\nu _{2}}\cdots {\nu _{L}}}=\int \frac{d^{D}k}{(2\pi )^{D}}\frac{%
k^{\nu _{1}}k^{\nu _{2}}\cdots k^{\nu _{L}}}{\left( n\cdot k\right) ^{N_{1}}}%
,\quad L\leq N_{1},  \label{FINAL_CONT}
\end{equation}%
each of them multiplied by a corresponding $k$-independent tensor. The
calculation of such integrals in dimensional regularization has been
previously discussed in references such as \cite{FRENKEL, MATSUKI,
LEIBBRANDT}. We briefly quote the results of Eqs.(2.5), (2.6) and (2.7) in
Ref. \cite{MATSUKI}. There, the author writes $I^{{\nu _{1}\nu _{2}}\cdots {%
\nu _{L}}}$ as%
\begin{equation}
I^{{\nu _{1}\nu _{2}}\cdots {\nu _{L}}}\sim \frac{\partial }{\partial n_{\nu
_{1}}}\frac{\partial }{\partial n_{\nu _{2}}}...\frac{\partial }{\partial
n_{\nu_{L}}}\int \frac{d^{D}k}{(2\pi )^{D}}\frac{1}{\left( n\cdot k\right)
^{(N_{1}-L)}},  \label{REDUCTION}
\end{equation}%
where the basic integral yields
\begin{equation}
\int \frac{d^{D}k}{(2\pi )^{D}}\frac{1}{\left( n\cdot k\right) ^{r}}%
=0,\;\;for\;r>0.
\end{equation}%
He considers separately the case $r=0,\;(L=N_{1})$, starting from
\begin{equation}
I^{{\nu _{1}\nu _{2}}...{\nu _{N_{1}}}}\sim \frac{\partial }{\partial n_{\nu
_{1}}}\frac{\partial }{\partial n_{\nu _{2}}}...\frac{\partial }{\partial
n_{\nu_{( N_{1}-1)} }}\int \frac{d^{D}k}{(2\pi )^{D}}\frac{k^{\nu
_{N_{1}}}}{\left( n\cdot k\right) }
\end{equation}%
and calculates the remaining integral assuming $\left( n\cdot k\right)
\rightarrow \left( n\cdot k\right) \pm i\epsilon $, obtaining also a zero
result.

In this way, using dimensional regularization, we have established that the
ghosts decouple when we write QED in the non-linear gauge $A_{\mu }A^{\mu
}=n^{2}M^{2}$, or equivalently in the axial gauge $n\cdot a=0\;$plus added
nonlinear interactions.

\section{Summary}

We prove that after imposing the Gauss law and current conservation as
initial conditions on the  ANM, the resulting theory  is equivalent  to QED
formulated in the non-linear gauge $A_{\mu }A^{\mu }-M^{2}=0$,  to all
orders in perturbation theory. The strategy is to write both  theories in
terms of fields describing the same degrees of freedom, which arise from the
same Lagrangian density thus yielding identical Feynman rules. In this way, the
perturbative calculations of any physical process in each model are
indistinguishable.

Our starting point in the fermionic ANM is Ref. \cite{Azatov-Chkareuli},
where the authors take the useful step  of introducing a further field
redefinition of $A_\mu$ in terms of the variables  $a_{\mu }$, which we  recall
in  Eq. (\ref{CHANGV}). In these variables, the constraint $A_{\mu }A^{\mu
}-M^{2}=0$ which defines the ANM, translates into the simpler form $%
\;n\cdot a=0$. This condition exhibits the fields $a_{\mu }$ as the pure GB
modes of the model, that are orthogonal to the direction $n_{\mu }$ of the
vacuum inducing the SLSB.  At this stage, the ANM is defined by the
Lagrangian density $\mathcal{L}_{\mathrm{ANM}}(a_{\mu },{\bar{\psi}},\psi )=\mathcal{%
L}_{\mathrm{QED}}(A_{\mu }(a_{\nu }),{\bar{\psi}},\psi )$, plus the
condition $n\cdot a=0$. This condition is a very convenient way of replacing
the initial nonlinear constraint because one can make explicit the
corresponding substitutions $a_{0}=0$, or $a_{3}=0$ at the end of the
calculation, thus allowing for a unified construction of the ANM in the
time-like and space-like cases, respectively. The requirement  $n\cdot a=0$
is effectively incorporated in the calculations through the propagator $D_{\mu \nu }^{\mathrm{ANM%
}}(k)$\ given in Eq. (\ref{FOTPROPANM}) and satisfying $n^{\mu }D_{\mu \nu }^{%
\mathrm{ANM}}=0$ together with \ $k^{\mu }D_{\mu \nu }^{\mathrm{ANM}}=0$\
for on-shell GB. Since the perturbative approach relies on the interaction
picture, the fields are quantized starting from the free Lagrangian density (\ref%
{LANMIP}) which, in particular, leads to the propagator $D_{\mu \nu }^{%
\mathrm{ANM}}(k)$. As emphasized in Ref. \cite{Nambu-Progr}, the on-shell
transversality of $D_{\mu \nu }^{\mathrm{ANM}}(k)$ guarantees that the
Gauss law is imposed, \`{a} la Dirac, upon the physical states. Moreover,
the free fermionic current is conserved, since it is  the Noether current arising
from the global phase invariance of the fermionic sector. In this way, the
perturbative approach ensures that the two additional requirements to be
imposed as initial conditions upon the ANM to recover gauge invariance are
indeed satisfied. The remaining  Feynman rules are then  obtained from the
expansion of the Lagrangian density $\mathcal{L}_{\mathrm{ANM}}(a_{\mu },{\bar{\psi}%
},\psi )=\mathcal{L}_{\mathrm{QED}}(A_{\mu }(a_{\nu }),{\bar{\psi}},\psi )\;$
in powers $a^{2}/M^{2}$.

Next we turn to QED and construct the BRST gauge fixed Lagrangian density, which
introduces the Faddeev-Popov ghosts ${\bar{c}},c$. We start from $\mathcal{L}%
_{\mathrm{QED}}(A_{\mu },{\bar{\psi}},\psi )$ choosing the gauge
$A_{\mu }A^{\mu }-M^{2}=0$,  but after we rewrite the gauge-fixed Lagrangian
density in terms of the same parameterization  (\ref{CHANGV}) used in the ANM,
observing that now the fields $a_{\mu }$ describe photons instead of
Goldstone bosons. The gauge fixing condition $\alpha (A_{\mu }A^{\mu
}-M^{2})^{2}$ translates into $4M^{2}\alpha (n\cdot a)^{2}$, which emerges
as the choice of an axial gauge, that nevertheless incorporates extra
Yang-Mills type interactions. By choosing the pure (homogeneous) axial
gauge, $\alpha \rightarrow \infty $, we arrive to the photon propagator $%
D_{\mu \nu }^{\mathrm{QED}}$ in Eq. (\ref{PROPLEIB}), which is identical to
the ANM propagator $D_{\mu \nu }^{\mathrm{ANM}}$ given in Eq. (\ref%
{FOTPROPANM}). At this stage, the relation between the two theories, written
in terms of the same variables and having the same Feynman rules for the
fields $a_{\mu },{\bar{\psi}},\psi $ can be summarized in the following
relation between their Lagrangian densities
\begin{equation}
\mathcal{L}_{\mathrm{ANM}}(a_{\mu },{\bar{\psi}},\psi )=\mathcal{L}_{\mathrm{%
QED}}(A_{\nu }(a_{\mu }),{\bar{\psi}},\psi )+\mathcal{L}_{\mathrm{GHOST}%
}(a_{\mu },{\bar{c}},c).
\end{equation}%
In this way, the last step to prove  the equivalence between them is to show
that the ghosts decouple to all orders in perturbation theory. Following a
method similar to that employed in Yang-Mills theories when investigating
the behavior of axial gauges, we prove in section 5 that this is indeed the
case. We make use of dimensional regularization and we consider the specific
photon-ghost interactions of the gauge fixed QED Lagrangian density.

Recapping, because the ghosts decouple, we have shown that the ANM,
supplemented by the initial conditions  mentioned earlier,  together with
QED written in the non-linear gauge $A_{\mu }A^{\mu }-M^{2}=0$, are
described by the same Lagrangian density $\mathcal{L}_{\mathrm{QED}}(A_{\nu }(a_{\mu
}),{\bar{\psi}},\psi )$. This yields to  identical Feynman rules for the propagators
and the interactions of the common fields $a_{\mu },{\bar{\psi}},\psi $,
thus making the two models identical in perturbation theory. The
perturbative calculations in Ref. \cite{Azatov-Chkareuli} constitute
detailed examples of this equivalence. Our general result agrees with the
Hamiltonian  approach discussed in Refs. \cite{Urru-Mont,NANM_ES_UR}. As
emphasized in such references, to prove the equivalence some additional
requirements had to be enforced upon the ANM as initial conditions, which
turned out to be valid for all times in virtue of ANM dynamics. These  were
current conservation and the imposition of the Gauss law upon the physical
states. As previously explained in the text, these conditions  are
fulfilled in the perturbative calculation, which is an expansion in terms
of fields in the interaction picture that satisfy free-field equations of
motion.

LFU is partly supported by the Project No. IN104815 from Direcci\'on General de Asuntos del Personal Acad\'emico (Universidad
Nacional Aut\'onoma de M\'exico).

\end{document}